\documentclass[sigconf, nonacm, authorsversion]{acmart}

\usepackage{booktabs} 
\usepackage{makecell}

\setcopyright{rightsretained}

\acmDOI{10.475/123_4}

\acmISBN{123-4567-24-567/08/06}

\acmConference[ARES 2019]{14th International ARES Conference on Availability, Reliability and Security}{August 2019}{Canterbury, UK}
\acmYear{2019}
\copyrightyear{2019}

\acmArticle{4}
\acmPrice{15.00}

\begin{document}
\title{Surfing the Web quicker than QUIC\\via a shared Address Validation}

\author{Erik Sy}
\affiliation{%
  \institution{University of Hamburg}
}


\begin{abstract}
QUIC is a performance-optimized secure transport protocol and a building block of the upcoming HTTP/3 standard.
To protect against denial-of-service attacks, QUIC servers need to validate the IP addresses claimed by their clients.
So far, the QUIC protocol conducts address validation for each hostname separately using validation tokens.
In this work, we review this practice and introduce a new QUIC transport parameter to allow a shared address validation across hostnames.
This parameter indicates to the client, that an issued validation token can be used to abbreviate the address validation when connecting to specific other hostnames.  
Based on trust-relations between real-world hostnames we evaluate the performance benefits of our proposal.
Our results suggest that a shared address validation saves a round-trip time on almost 60\% of the required handshakes to different hosts during the first loading of an average website. 
Assuming a typical transatlantic connection with a round-trip time of 90~ms.
We find that deploying our proposal reduces the delay overhead to establish all required connections for an average website by 142.2~ms.  
\end{abstract}

%
%



\maketitle

\section{Introduction}
The world wide web is closely tied to the HTTP network protocol.
The upcoming version HTTP/3 will replace the traditional TLS over TCP stack with the new QUIC protocol~\cite{ietf-quic-http-19}.
Thus, it is likely that QUIC will be widely adopted on the web within the next couple of years.
Further potential use cases include, but are not limited to the Domain Name System (DNS)~\cite{huitema-quic-dnsoquic-06}.

A principal goal of the QUIC protocol is to reduce the delay overhead required to establish secure connections.
To achieve this goal QUIC provides the feature of a zero round-trip time handshake that conducts the transport and the cryptographic connection establishment at the same time.
However, this feature is only applicable when reconnecting to a QUIC server as it requires cached state of a prior connection.
The first connection to a hostname requires up to two additional round-trips.
One additional round-trip is caused by the cryptographic connection establishment and the other by the transport handshake.

In this work, we are addressing the performance limitations caused by this additional round-trip of the transport handshake that is used to validate the client's source address.
To illustrate the shortcomings of QUIC's current design, we assume that a client connects sequentially to the hostnames \textit{example.com} and \textit{www.example.com}.
Furthermore, we assume both of these hostnames are operated by the same entity and a strict address validation is deployed by the responding QUIC servers.
We find that the default QUIC behavior causes per connection one additional round-trip for the validation of the client's address.
As a result of this process, the client's address has been validated at the expense of two additional round-trips by the same entity via different hostnames.

We propose a performance-optimized approach that allows to reuse address validation tokens across hostnames that have a trust-relation to each other.
The client connects to the first hostname with an additional round-trip to validate the client's address and receives a validation token.
The second handshake to the other hostname uses the validation token received during the first connection to pass the address validation without an additional round-trip.
This approach saves in total a round-trip time compared to the default QUIC behavior

In summary, this paper makes the following contributions:

\begin{itemize}

\item We propose a new transport parameter for the QUIC protocol that enables a shared address validation across different hostnames.
\item We demonstrate the performance gains yielded by our proposal for loading popular websites. Our results indicate that deploying the introduced transport parameter saves a round-trip time on almost 60\% of the required connection establishments to different hosts during the first loading of an average website.

\end{itemize}

The remainder of this paper is structured as follows: Section~\ref{sec:Problem} describes the performance problem of the QUIC transport protocol that we aim to solve. 
Section~\ref{sec:Delegation} summarizes the proposed shared address validation and evaluation results are presented in Section~\ref{sec:Evaluation}.
Related work is reviewed in Section~\ref{sec:Related}, and Section~\ref{sec:Conclusion} concludes the paper.

\section{Problem Statement} \label{sec:Problem}

\begin{figure*}[htpb]
\centering
\begin{minipage}{.3\linewidth}
    \includegraphics[width=\linewidth]{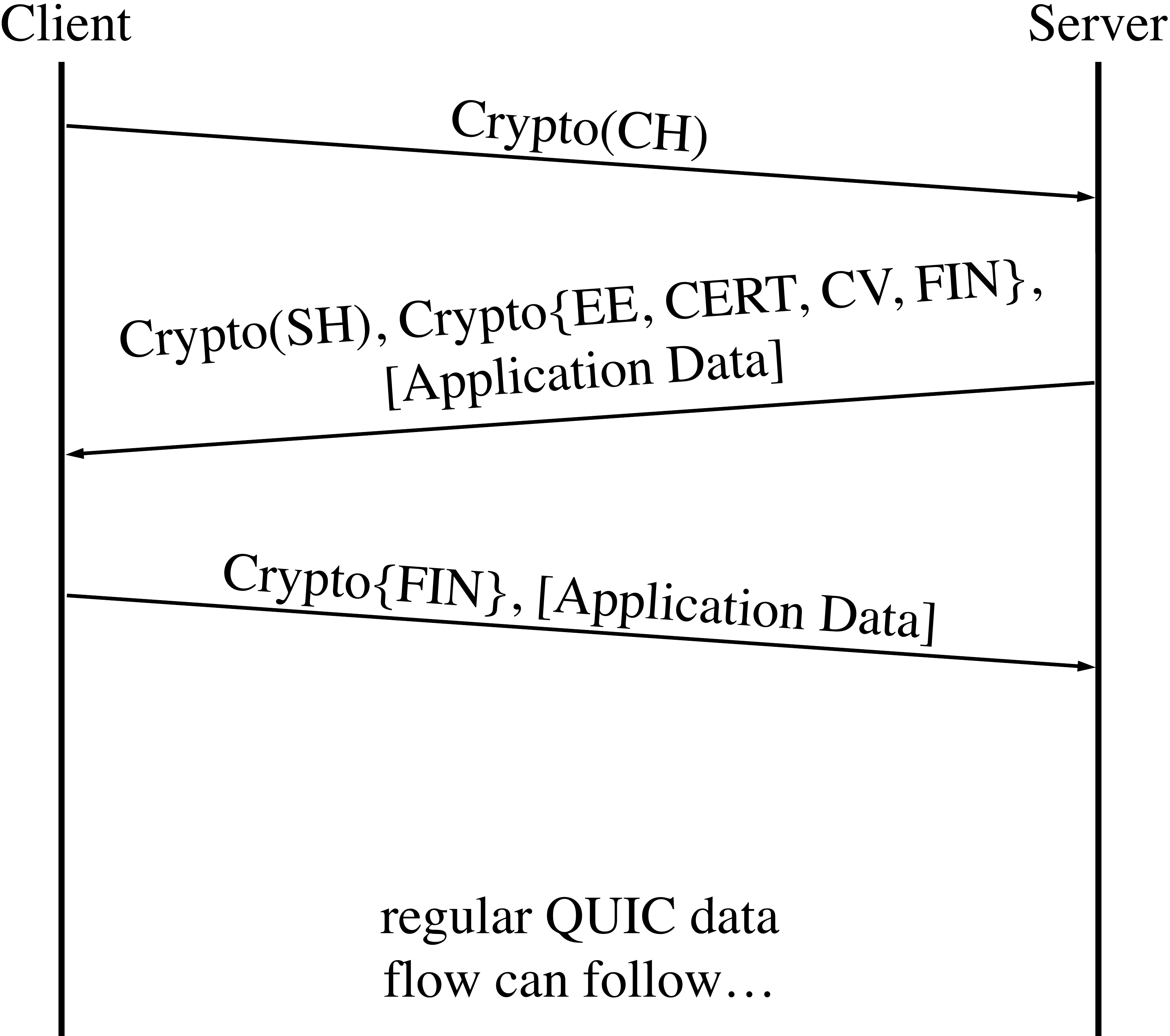}
    \caption*{a) Initial handshake}
\end{minipage}
\hfill
\begin{minipage}{.3\linewidth}
    \includegraphics[width=\linewidth]{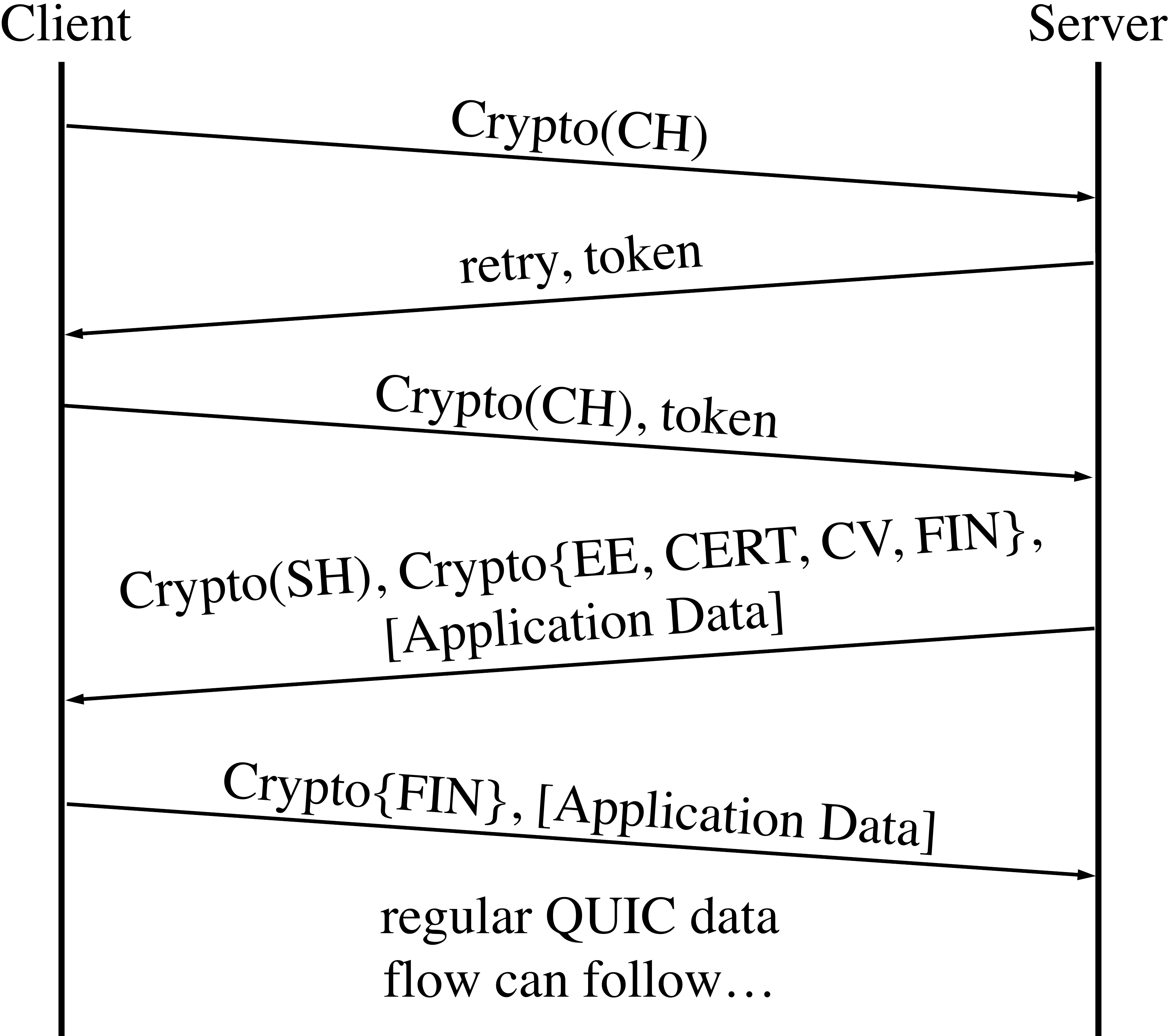}
    \caption*{b) Initial handshake with retry}
\end{minipage}
\hfill
\begin{minipage}{.3\linewidth}
    \includegraphics[width=\linewidth]{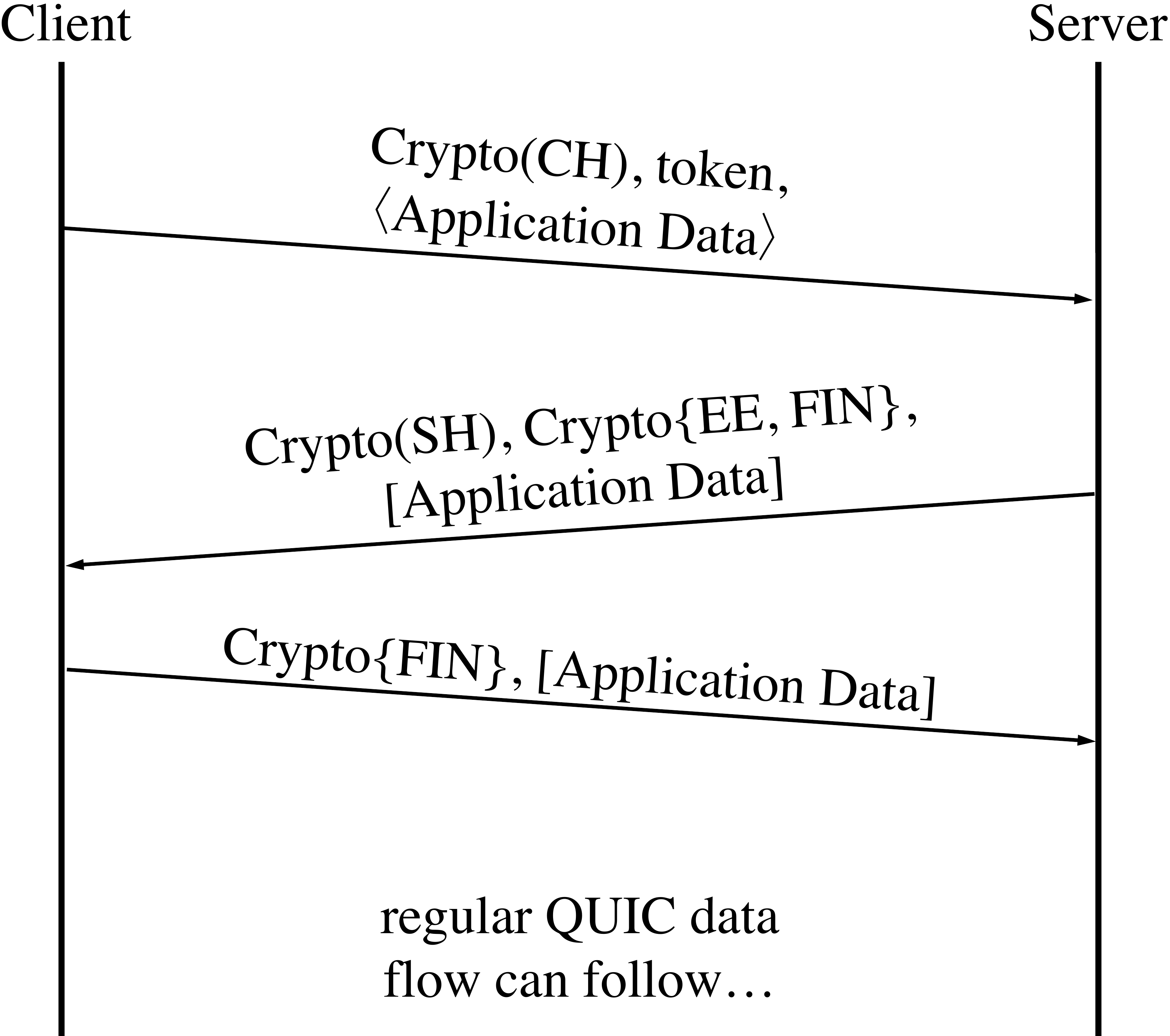}
    \caption*{c) 0-RTT handshake}
\end{minipage}
  \caption{Handshakes in IETF QUIC protocol, where the brackets indicate different levels of encryption.
  Round brackets indicate no encryption and curved brackets denote encryption based on the handshake traffic secret.
  Square brackets signal encryption using the application traffic secret and angle brackets indicate the use of the early traffic secret for encryption.}
  \label{fig:QUIC_overview}
  \vskip -12pt
\end{figure*}

In this section, we briefly review the connection establishment of the IETF QUIC protocol~\cite{ietf-quic-transport-19}.
Subsequently, we describe the problem that we aim to solve and our threat model.
\subsection{Connection establishment with IETF QUIC}

Google developed and deployed the initial design of the QUIC protocol~\cite{langley2017quic}, which we will refer to as Google QUIC (gQUIC).
In 2016, the Internet Engineering Task Force (IETF) started to standardize QUIC~\cite{ietf-quic-transport-19}, which is still a work in progress.
The IETF's draft of QUIC differs significantly from gQUIC, e. g., it uses TLS~1.3~\cite{rfc8446} to conduct the cryptographic handshake.
However, both QUIC variants aim to improve the performance of HTTPS traffic by conducting the cryptographic and the transport handshake concurrently, while the widespread approach using TLS over TCP conducts these handshakes sequentially.
Figure~\ref{fig:QUIC_overview} shows a schematic of QUIC's connection establishment. 

\paragraph{Initial handshake with and without retry}
As shown in Figure~\ref{fig:QUIC_overview}~a), the client starts with the ClientHello (CH) message containing lists of supported cipher suites, protocol versions, TLS extensions, and public keys suitable for key exchange.
All messages that are part of the cryptographic connection establishment follow the TLS~1.3~\cite{rfc8446} protocol and are emphasized with \textit{Crypto} in Figure~\ref{fig:QUIC_overview}.
The server responds with its unencrypted ServerHello message, which is indicated by round brackets in Figure~\ref{fig:QUIC_overview}.
Subsequently, the server computes the handshake traffic secret and sends the Encrypted Extensions (EE), the server's certificate (CERT), the Certificate Verify (CV), and the handshake finished (FIN) messages encrypted with this secret, shown as curved brackets in Figure~\ref{fig:QUIC_overview}.
In the CV message the server provides a fresh proof for its ownership of the certificate's private key.
Whereas FIN messages signal a successful handshake and contain hashes of the exchanged handshake messages to verify that both peers observed the same messages.
The server can now calculate the application traffic secret and may send encrypted application data.
This encryption type is indicated with square brackets in Figure~\ref{fig:QUIC_overview}.

Upon receiving the ServerHello message, the client computes the handshake traffic secret.
This enables the client to decrypt the received EE, CERT, CV, and FIN messages.
The client can now authenticate the server's identity based on the provided CERT and CV messages.
Then, the client validates the hashes contained in the server's FIN message and calculates the application traffic secret.
Finally, the client responds with its own FIN message and may send encrypted application data to the server.

After a connection is successfully established, the server can provide the client with TLS resumption tickets and address validation tokens that can be used on subsequent connections.
Resumption tickets allow resuming previous connections via abbreviated handshakes that decrease the delay overhead and save expensive cryptographic computations during connection establishment.
A validation token is an encrypted and authenticated data block which is opaque to the client.
It usually contains the client's visible IP address as seen by the server.
If the client provides such a token during a subsequent connection establishment, this allows the server to compare the client's claimed IP address with the previously observed clients' IP address in the token.

Depending on its configuration, the server may per default or dynamically decide to strictly validate the client's IP address before proceeding with the cryptographic handshake.
Figure~\ref{fig:QUIC_overview}~b) shows an initial handshake that validates the client's source address with an additional round-trip.
Here, the server sends a retry message and an address validation token to the IP address claimed with the client's first message.
To proof the ownership of the claimed IP address, the client is required to provide the received token along with its ClientHello message.
Upon receiving the token from the client, the server validates it, which involves a comparison of the IP address stored in the token with the one claimed by the client.   
If the token is valid, the client-server pair will proceed with a standard initial connection establishment starting from the ServerHello message (see Figure~\ref{fig:QUIC_overview}~a).

\paragraph{0-RTT connection establishment}
Furthermore, the QUIC protocol provides zero round-trip time (0-RTT) connection establishments as shown in Figure~\ref{fig:QUIC_overview}~c).
Here, the client can send data encrypted with the early traffic secret without waiting for a response from the server using TLS resumption.
Thus, the client requires a previously retrieved resumption ticket and a pre-shared key (PSK) to encrypt these early data and to signal the used PSK to the server. 
Optionally, the client can provide an address validation token as shown in Figure~\ref{fig:QUIC_overview}~c) to anticipate a server's retry request.

Upon receiving these messages, the server starts to validate the client's token and IP address.
In case of a positive validation result, the server begins with its own cryptographic operations.
To derive the early traffic secret, the server uses also information provided in the \emph{pre\_shared\_key} extension of the ClientHello message.
Assuming, that the server can successfully decrypt the provided early data, it will signal this with the \emph{pre\_shared\_key} extension in the ServerHello message.
After sending the ServerHello message, the server will derive the handshake traffic secret and use it for encrypted transmission of the EE and FIN messages.
Note, that this handshake does not require the server to provide a CERT and CV message to authenticate its claimed identity.
Instead, the peers authenticate each other by successfully resuming the previous TLS session using the pre-shared key.
This abbreviated authentication during resumed connections significantly decreases the delay and saves expensive cryptographic operations.
Subsequently, the server can derive its application data secret and respond with encrypted application data to the client's early data.

Upon receiving the server's messages, the client will derive the handshake traffic secret and decrypt the server's EE and FIN message.
Then, the client reviews the validation data contained in the server's FIN message to validate, that the exchanged messages have not been tampered with during transit. 
If the client did not determine a modification of the received data, it will provide the server its own FIN message. 
Finally, the client can derive its application traffic secret and the regular data flow follows.

Note, that application data encrypted with the early traffic secret does not provide forward-secrecy and might be replayed in other connections.
For a comprehensive discussion on the weaker security guarantees of early data, we refer readers to RFC~8446~\cite{rfc8446}.

\subsection{Performance limitation of address validation}

A QUIC server can validate the IP address claimed by a client by sending it a retry message with a token.
Only, if the client returns this issued token, the server proceeds with the cryptographic handshake.
This practice protects the server from spending expensive cryptographic operations on malicious handshake attempts claiming an illegitimate IP address. 
As a drawback, this IP spoofing defense increases the delay overhead of the connection establishment by a round-trip time.

To avoid this additional delay during the establishment of a new connection, a client can present a token from a previous connection to the same hostname along with the ClientHello message.
If the server accepts this token as valid for the IP address claimed by the client, then it will directly proceed with the cryptographic connection establishment.
This practice requires the client to retrieve a token in a previous QUIC connection to the same hostname.

QUIC does not consider using a retrieved address validation token to connect to new hostnames not matching the connection that has been used to retrieve the token.
As a result, connections to every fresh hostname cannot present an address validation token leading to the described additional delay overhead if the server enforces address validation.
This is a performance limitation if trust-relations between different hostnames are available.
To illustrate this limitation, we assume a trust-relation between the hostnames \textit{example.com} and \textit{www.example.com}, which mutually accept their issued address validation tokens.
Assuming the same network latency to both hostnames, a client establishing fresh connections to both hostnames experiences an additional delay overhead of twice the round-trip time due to address validation.
However, if the client sequentially connects to these hostnames and reuses a token obtained in the first connection to establish the latter connection, this bisects the additional delay overhead to a single round-trip time.

To put this into perspective, a typical latency in North America is <45\,ms and <90\,ms for transatlantic connections~\cite{Verizon}.
However, several regions in the world exist which suffer from high network delays, often exceeding 300\,ms~\cite{formoso2018deep}.
In total, this example illustrates that using address validation tokens across different hostnames can provide significant performance gains, especially for connections experiencing high latencies.

\subsection{Threat Model}

To clarify the security aspects of the described problem, we define our threat model in the following.

The considered adversary is able to spoof the source address of its packets.
However, we assume the address validation tokens to be cryptographically secure, thus attackers cannot generate valid tokens.
In total, our adversary affects the security objective of availability.
By spoofing the IP address of a victim's endpoint, the attacker causes the QUIC server to send its response to the victim, which is also known as a reflection attack.
Moreover, the attacker can also directly attempt to exhaust the resources of the server by requesting connections from various spoofed IP addresses.

\section{Shared address validation across hostnames} \label{sec:Delegation}

This section describes our approach of a shared address validation across hostnames for the QUIC protocol.
For that, we introduce the new transport parameter \textit{validation\_group} that aims to enable a shared address validation across hostnames.

Deviating from the standard initial handshake, the server unilaterally includes the \textit{validation\_group} value in its list of transport parameters. 
In detail, the \textit{validation\_group} presents a one-bit value, which is set to one if this feature is supported and zero otherwise.

The client finds the declared \textit{validation\_group} value in the server's EE message.
If this value is set to zero, then the client reasons that the server does not support this feature and proceeds with its default behavior.

In case the \textit{validation\_group} value is set to one, the client concludes that the server supports a shared address validation across the hostnames, for which the presented TLS certificate is valid.
We define a validation group to be a list of hostnames for which a single TLS certificate is valid and that mutually support address validation using tokens issued by any member of the same group.
To support a shared address validation, the client associates received tokens during that connection with the validation group at hand.

To open a new connection to any member of a validation group, the client can use cached tokens associated to the same group.
Tokens received during such a new connection must be associated with the same validation group.

The QUIC server does not provide a TLS certificate during a resumed connection establishment.
In this case, the tokens received during the connection should be associated with the TLS certificate of the original connection, for which the server's identity was authenticated using a TLS certificate.

To mitigate tracking via address validation tokens~\cite{sy2019quic}, a client-side expiration mechanism is required for them.
The lifetime of these tokens presents a performance versus privacy trade-off, where shorter lifetimes lead to better privacy protection.
Based on an analysis of characteristic Internet traffic, the authors of~\cite{sy2018tracking} recommend a lifetime of ten minutes to address this trade-off in web browsers.

To support a shared address validation across hostnames, the cryptographic secret used to generate and decrypt the tokens needs to be shared across the servers serving these hostnames.
Note, that address validation tokens are designed for single-use only.
If a member of a validation group is not properly configured and fails to validate a legitimate token, then this process uses up a cached token of the respective validation group.
In the worst case, this behavior can lead to reduced performance compared to the status quo.
For example, if the client needs to respond to a retry message of a server because it used up its cached tokens on not properly configured servers.

\section{Evaluation} \label{sec:Evaluation}

In this section, we evaluate the performance benefit of our proposal for web browsing.
For this purpose, we simulate shared address validation across hostnames for the Alexa Top 1K Sites~\cite{Alexa}.
We start by describing our assumptions for this evaluation.
Subsequently, we introduce the analyzed data set and summarize our results.

\subsection{Assumptions}
To simulate the loading behavior of popular websites, we assume that each established connection enforces a strict address validation.
This assumption describes the current practice of TCP Fast Open~\cite{rfc7413} and gQUIC~\cite{langley2017quic}.
However, IETF QUIC provides an operation mode with a relaxed address validation.
Nonetheless, we believe the strict address validation to become the most popular operation mode of QUIC servers.
As the draft on IETF QUIC is still in an early stage, we are not aware of online services deploying IETF QUIC at a production level.
Thus, we cannot determine the configuration of real-world deployments of IETF QUIC to assess the accuracy of this assumption at the moment. 

Moreover, this evaluation assumes that the investigated websites deploy the QUIC protocol to support their HTTPS connection establishment.
To justify this assumption, we point out that numerous browser vendors and content delivery networks contribute to the draft of the QUIC protocol~\cite{ietf-quic-transport-19}
and intend to deploy this protocol in their products.
Furthermore, the upcoming HTTP/3 uses the QUIC protocol~\cite{ietf-quic-http-19} by default, thus it will be deployed to serve websites.
In total, it seems likely that the QUIC protocol will be widely deployed on the Internet within the next couple of years.

In addition, to simulate our proposal for popular websites we require information on real-world hostnames that trust each other with regard to the availability of their servers.
To approximate this information, we assume that hostnames trusting each other with respect to the confidentiality of their communications are also willing to have a trust-relation concerning the availability of their servers.
We define that a trust-relation with respect to confidentiality exists, if hostnames share a secret cryptographic state with each other such as the private key of their TLS certificate or they enable the resumption of TLS sessions across their hostnames.
From our perspective, it seems reasonable that such closely cooperating hostnames, which are often operated by the same entity, are willing to trust each other in terms of the validation of their clients' IP addresses.

\subsection{Data set}

This paper uses a data set on trust-relations of the Alexa Top 1K Sites, that is described and evaluated in~\cite{sy2019enhanced}.
The data set aggregates a scan of the Alexa Top 1K Sites~\cite{Alexa} performed on the 8th of November 2018.
In total, the scan successfully retrieved the domain trees of 839 websites.
A domain tree describes an overview on the sequence of established connections to different hostnames during the retrieval of a website (see Figure~\ref{fig:Domain_tree}).
Furthermore, the data set collected real-world TLS trust-relations between the different hostnames within each domain tree.
For that, the data set defines a TLS trust-relation between two hostnames if they either share the same TLS certificate or enable the resumption of TLS sessions between their hostnames.

\subsection{Results}

In this section, we demonstrate the performance benefits of our proposal for the first loading of popular websites, respectively.
First, we investigate the amount of connections necessary for the retrieval of an average website, that can each save a round-trip time during the connection establishment by deploying shared address validation.
Subsequently, we observe that most websites require a client to sequentially establish connections to different hosts.
Based on this practice, we analyze our proposal by measuring the aggregated delay overhead for establishing all QUIC connections necessary for the retrieval of the respective website successively.
This evaluation of successive retrievals allows to approximate the benefit of our proposal for the loading time of a website.

\begin{figure}
\centering
\includegraphics[width=0.47 \textwidth]{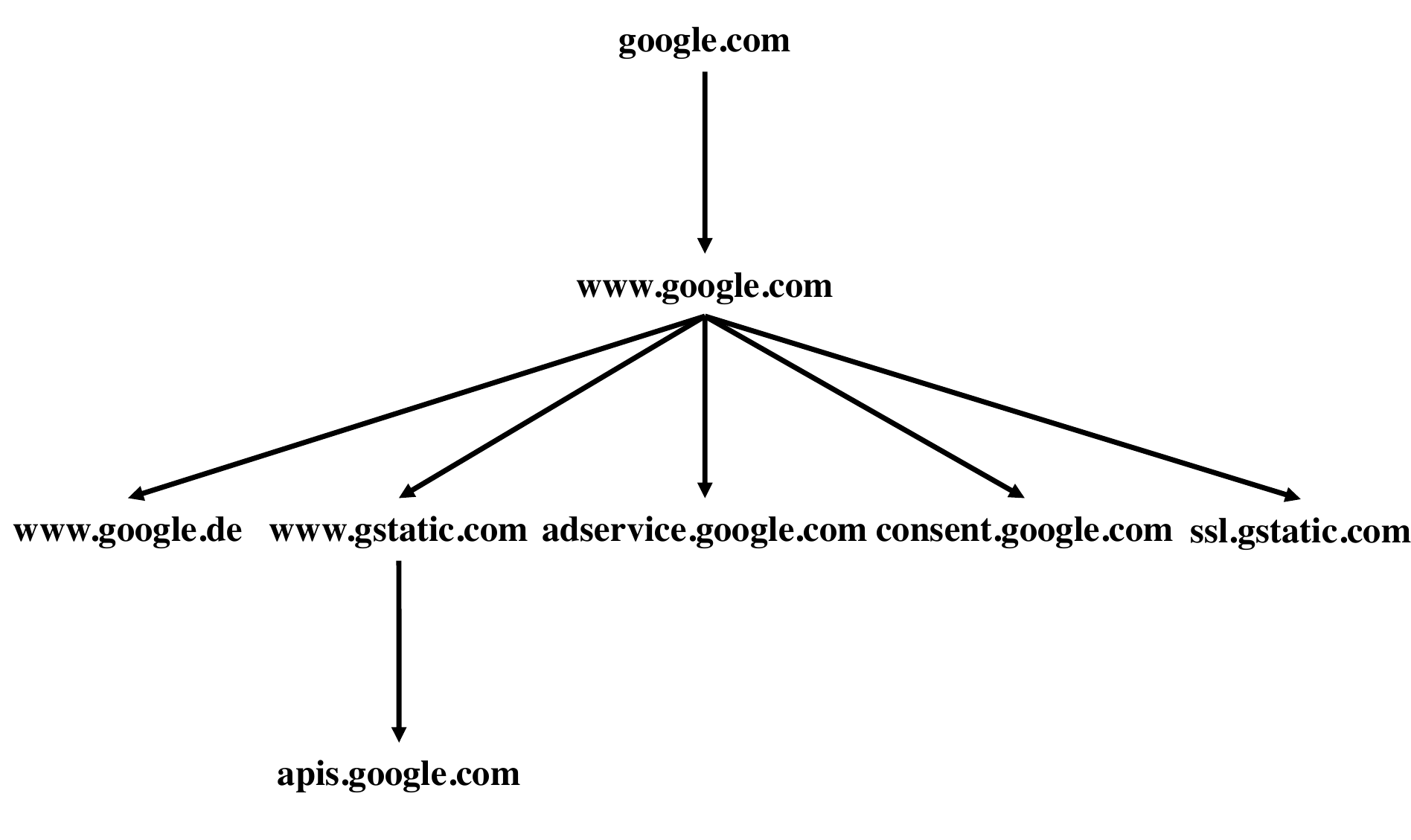}
\caption{Domain tree of the website google.com. }
\label{fig:Domain_tree}
\end{figure}

\paragraph{Share of abbreviated handshakes}

\begin{figure}[tbp]
\centering
\includegraphics[width=0.47 \textwidth]{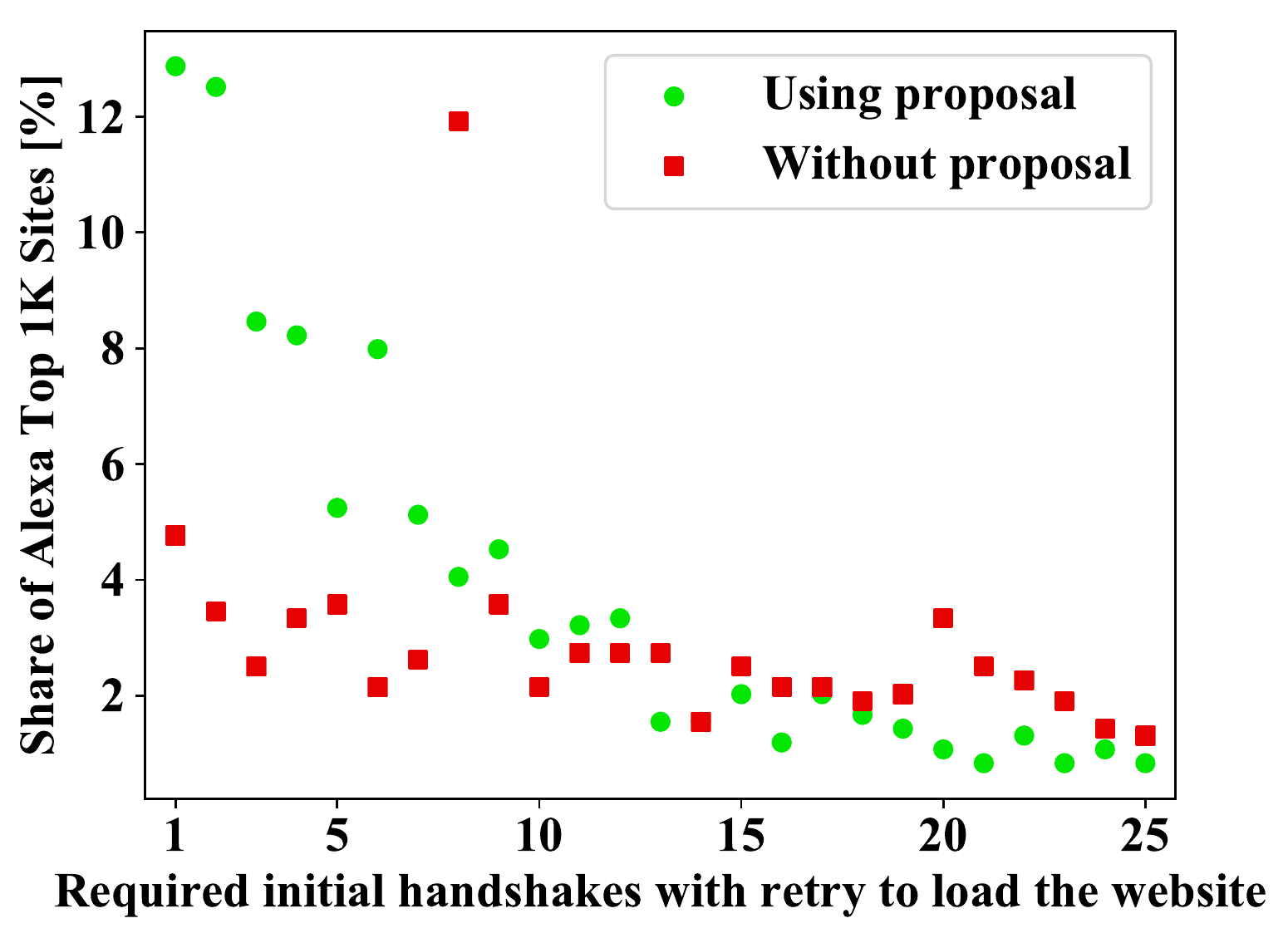}
\caption{This plot shows the share of Alexa Top~1K Sites in dependence on the number of required initial handshakes with retry to retrieve the website. The green circles mark the values considering the introduced shared address validation, while the red squares plot the current default. Note, that this plot is cut off at 25 handshakes.}
\label{fig:trustgroup_number}
\end{figure} 

The analyzed data set~\cite{sy2019enhanced} indicates, that TLS trust-relations are a common practice on the web.
In total, the average Alexa Top 1K Site requires the client to establish 20.24 encrypted connections to different hostnames.
Figure~\ref{fig:trustgroup_number} plots the distribution of the share of these websites over the number of required initial handshakes with retry for their retrieval.
The plot marked by the red squares indicates that 95.2\% of these websites require more than a single secure connection.
Note, that 73.3\% of the investigated sites can be loaded with at most 25 initial handshakes with retry.
Thus, we cut off this plot after 25 of these handshakes for reasons of clarity.

The plot marked with green circles presents the results of our simulation using a shared address validation in case of a trust-relation between the corresponding hostnames.
Thus, it converts an initial handshake with retry to an initial handshake where the client presents an address validation token received from a member of the same validation group.
As a result, each of these conversions reduces the delay overhead of the specific connection establishment by a round-trip.
Figure~\ref{fig:trustgroup_number} shows that the use of a shared address validation shifts the distribution of the Alexa Top 1K Sites towards less required initial handshakes with retry.
For example, the share of sites requiring a single initial handshake with retry increased from 4.8\% to 12.9\% by using our proposal.
Furthermore, we find that 42.1\% of the investigated websites can be retrieved with less than five initial handshakes with retry by using the introduced proposal compared to 14.1\% without deploying the proposal. 
In total, we find that 96.0\% of the Alexa Top 1K Sites can be retrieved with at most 25 initial handshakes with retry by deploying a shared address validation. 

 \begin{table}[htbp]
\caption{Mean number of required initial QUIC handshakes with retry to different hostnames to download a website of the Alexa Top~1K list for the first time in absolute and relative numbers.}\label{tab:fullhandshake_cross_host}
 \centering
 \begin{tabular}{ccc}
 \toprule
\makecell{Without shared\\address validation} & \makecell{With shared\\address validation}& \makecell{Savings}\\
 \midrule
20.24 & 8.35 & 11.89 \\
100.0\% & 41.25\% & 58.75\%\\
 \bottomrule
 \end{tabular}
 \end{table}

Table~\ref{tab:fullhandshake_cross_host} provides an overview of the performance benefit of our proposal for the average Alexa Top 1K Site.
Our results suggest, that a shared address validation reduces the number of initial handshakes with retry from 20.24 to 8.35 for the first visit of an average website.
Thus, 11.89 initial QUIC connections can be converted to use an address validation token received from a member of the same validation group.
In total, the proposed practice saves for 58.75\% of the established QUIC connections a round-trip time.
This yields a reduction of 11.89 round-trips during the establishment of the required connections.

\paragraph{Delay overhead}

The studied data set~\cite{sy2019enhanced} provides insights into the sequence of established connections and which retrieved resources triggered the establishment of additional connections.
Figure~\ref{fig:Domain_tree} provides the domain tree of \textit{google.com} that marks initiated connections to different hostnames with arrows.
We observe that the longest path within the domain tree requires four sequential connection establishments via the hostnames \textit{google.com}, \textit{www.google.com}, \textit{www.gstatic.com}, and \textit{apis.google.com}.
Thus, the website retrieval of \textit{google.com} is impacted by about four times the delay overhead of a single connection establishment. 

\begin{figure}[tbp]
\centering
\includegraphics[width=0.47 \textwidth]{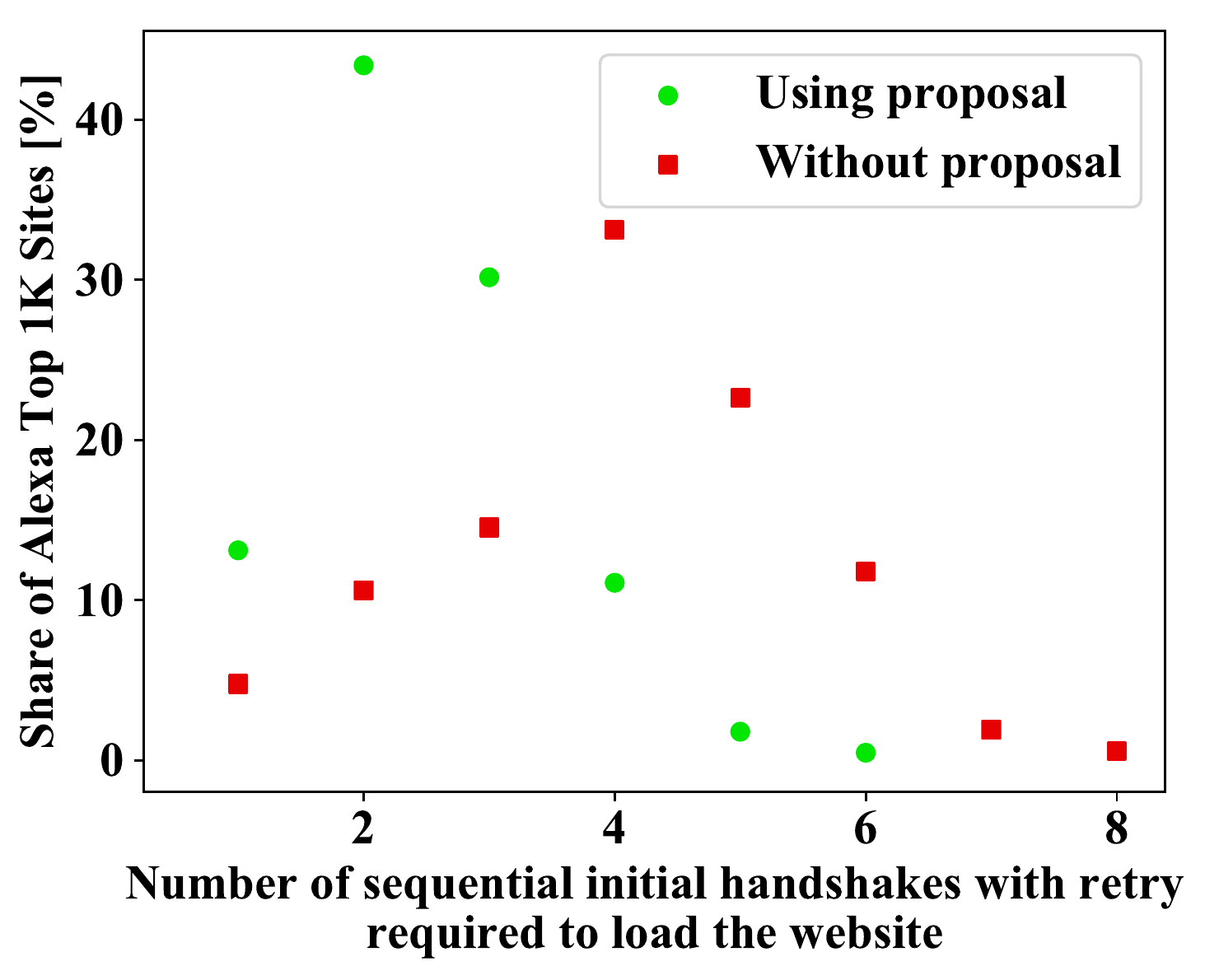}
\caption{This plot shows the share of Alexa Top~1K Sites over the number of required sequential initial QUIC connections with retry to load the respective website. The green circles represent the values considering  the introduced shared address validation, while the red squares plot the current default.}
\label{fig:longest_fullhandshake_cross_host_path}
\end{figure}

Figure~\ref{fig:longest_fullhandshake_cross_host_path} plots the distribution of Alexa Top 1K Sites over the length of their longest paths of initial handshakes with retry.
The plot using the red squares indicates the current status quo.
We observe, that the Alexa Top 1K Sites require no more than eight sequential connections for their retrieval.
The single most popular configuration requires four sequential connections and is used by 33.1\% of the Alexa Top 1K Sites.
In total, we find that 63.0\% of the Alexa Top 1K Sites can be retrieved with four or less sequential connections.

The plot marked with green dots provides the results for deploying a shared address validation.
In our simulation, we convert if applicable an initial handshake with retry to an initial handshake, where the client presents an address validation token obtained from another member of the same validation group.
Each of these converted handshakes saves a round-trip time during the corresponding connection establishment.
Note, that our evaluation repeats the computation of the longest paths of initial handshakes with retry after simulating the shared address validation.
Thus, the longest paths with and without using the introduced shared address validation can deviate from each other.

We find, that the deployment of our proposal leads to a significant reduction of necessary initial handshakes with retry for the Alexa Top 1K Sites.
For example, using a shared address validation reduces the share of websites requiring more than five sequential connections from 37.0\% to 2.3\%.
Furthermore, the number of websites requiring less than three connections increased from 15.4\% to 56.5\% when using our proposal.

Table~\ref{tab:full-handshake-request-path} provides an overview of the average benefit of our proposal.
Our results indicate, that the average Alexa Top 1K Site requires 4.04 sequential initial handshakes with retry.
The proposed shared address validation reduces this value to 2.46.
In total, our proposal saves 1.58 times the round-trip time until the last connection required for loading an average website is established.
This presents a reduction of the longest path of initial handshakes with retry by 39.1\% for the average website.
Assuming a transatlantic connection with a round-trip time of 90~ms~\cite{Verizon}, our proposal reduces the delay overhead for establishing all required QUIC connections by 142.2~ms.

 \begin{table}[htbp]
   \caption{Mean length of the longest path of required initial QUIC handshakes with retry to different hostnames to retrieve an average website of the Alexa Top~1K list for the first time in absolute and relative numbers. }\label{tab:full-handshake-request-path}
 \centering
 \begin{tabular}{ccc}
 \toprule
\makecell{Without shared\\address validation} & \makecell{With shared\\address validation}& \makecell{Savings}\\
 \midrule
4.04 & 2.46 & 1.58 \\
100.0\% & 60.89\% & 39.10\%\\
 \bottomrule
 \end{tabular}
 \end{table}
 
\section{Related Work} \label{sec:Related}

To the best of our knowledge, we are the first to investigate the benefit of a shared address validation across hostnames for transport handshakes.
Prior work includes the current draft of the QUIC transport protocol, as it is designed to support a shared address validation across servers serving the same hostname.

Furthermore, related work~\cite{sy2019enhanced} investigated trust-relations within the domain trees of the Alexa Top 1K Sites.
However, that work focused on performance improvements for the cryptographic handshake when using TLS session resumption across hostnames.

Moreover, HTTP version 2 (HTTP/2)~\cite{rfc7540} can be considered as related work as it allows to reuse TLS connections across different hostnames.
This approach in HTTP/2 aims to reduce the number of established connections to yield performance gains.
However, our proposal reduces the delay overhead of the connection establishment itself.

\section{Conclusions}\label{sec:Conclusion}

This work proposes a new transport parameter for the QUIC protocol to support clients to use address validation tokens across hostnames.
Our evaluation demonstrates, that such a shared address validation significantly reduces the delay overhead of QUIC's connection establishment on the real-world web.

\bibliographystyle{ACM-Reference-Format}
\bibliography{sample-bibliography}

\end{document}